\documentclass[aps,prl,twocolumn,showpacs]{revtex4-1}
\usepackage{epsfig,array}
\usepackage{amsbsy,latexsym}
\usepackage{amsmath}
\usepackage{amssymb, mathrsfs}
\usepackage[mathscr]{eucal}
\usepackage[active]{srcltx}

\def\<{\langle}\def\>{\rangle}
\def\Tr{\operatorname{Tr}}\def\:{\hbox{\bf :}}

\def\dag{\dagger}
\def\leq{\leqslant}

\def\Lin{\mathcal{L}}

\def\qed{$\,\blacksquare$\par}

\newtheorem{Def}{Definition}

\newtheorem{theorem}{Theorem}
\newtheorem{conjecture}{Conjecture}
\def\Proof{\medskip\par\noindent{\bf Proof. }}

\newcommand{\Ket}[1]{| #1 \rangle \! \rangle}
\newcommand{\Bra}[1]{\langle \! \langle #1 |}

\def\rA{{\rm A}}\def\rB{{\rm B}}\def\rE{{\rm E}}\def\rJ{{\rm J}}
\def\rX{{\rm X}}\def\rW{{\rm W}}
\def\Lin{{\mathcal L}}
\usepackage[matrix,frame,arrow]{xy}
\usepackage{amsmath}
\newcommand{\bra}[1]{\left\langle{#1}\right\vert}
\newcommand{\ket}[1]{\left\vert{#1}\right\rangle}
\newcommand{\qw}[1][-1]{\ar @{-} [0,#1]}
\newcommand{\qwx}[1][-1]{\ar @{-} [#1,0]}
\newcommand{\cw}[1][-1]{\ar @{=} [0,#1]}
\newcommand{\cwx}[1][-1]{\ar @{=} [#1,0]}
\newcommand{\gate}[1]{*{\xy *+<.6em>{#1};p\save+LU;+RU **\dir{-}\restore\save+RU;+RD **\dir{-}\restore\save+RD;+LD **\dir{-}\restore\POS+LD;+LU **\dir{-}\endxy} \qw}

\newcommand{\measureD}[1]{*{\xy*+=+<.5em>{\vphantom{\rule{0em}{.1em}#1}}*\cir{r_l};p\save*!R{#1} \restore\save+UC;+UC-<.5em,0em>*!R{\hphantom{#1}}+L **\dir{-} \restore\save+DC;+DC-<.5em,0em>*!R{\hphantom{#1}}+L **\dir{-} \restore\POS+UC-<.5em,0em>*!R{\hphantom{#1}}+L;+DC-<.5em,0em>*!R{\hphantom{#1}}+L **\dir{-} \endxy} \qw}

\newcommand{\multimeasureD}[2]{*+<1em,.9em>{\hphantom{#2}}\save[0,0].[#1,0];p\save !C *{#2},p+LU+<0em,0em>;+RU+<-.8em,0em> **\dir{-}\restore\save +LD;+LU **\dir{-}\restore\save +LD;+RD-<.8em,0em> **\dir{-} \restore\save +RD+<0em,.8em>;+RU-<0em,.8em> **\dir{-} \restore \POS !UR*!UR{\cir<.9em>{r_d}};!DR*!DR{\cir<.9em>{d_l}}\restore \qw}
\newcommand{\control}{*!<0em,.025em>-=-{\bullet}}
\newcommand{\controlo}{*-<.21em,.21em>{\xy *=<.59em>!<0em,-.02em>[o][F]{}\POS!C\endxy}}
\newcommand{\ctrl}[1]{\control \qwx[#1] \qw}

\newcommand{\qswap}{*=<0em>{\times} \qw}
\newcommand{\multigate}[2]{*+<1em,.9em>{\hphantom{#2}} \qw \POS[0,0].[#1,0];p !C *{#2},p \save+LU;+RU **\dir{-}\restore\save+RU;+RD **\dir{-}\restore\save+RD;+LD **\dir{-}\restore\save+LD;+LU **\dir{-}\restore}
\newcommand{\ghost}[1]{*+<1em,.9em>{\hphantom{#1}} \qw}

\newcommand{\gategroup}[6]{\POS"#1,#2"."#3,#2"."#1,#4"."#3,#4"!C*+<#5>\frm{#6}}
\newcommand{\rstick}[1]{*!L!<-.5em,0em>=<0em>{#1}}

\newcommand{\ustick}[1]{*!D!<0em,-.5em>=<0em>{#1}}

\newcommand{\Qcircuit}[1][0em]{\xymatrix @*[o] @*=<#1>}

\newcommand{\pureghost}[1]{*+<1em,.9em>{\hphantom{#1}}}
\newcommand{\multiprepareC}[2]{*+<1em,.9em>{\hphantom{#2}}\save[0,0].[#1,0];p\save !C
  *{#2},p+RU+<0em,0em>;+LU+<+.8em,0em> **\dir{-}\restore\save +RD;+RU **\dir{-}\restore\save
  +RD;+LD+<.8em,0em> **\dir{-} \restore\save +LD+<0em,.8em>;+LU-<0em,.8em> **\dir{-} \restore \POS
  !UL*!UL{\cir<.9em>{u_r}};!DL*!DL{\cir<.9em>{l_u}}\restore}
\newcommand{\prepareC}[1]{*{\xy*+=+<.5em>{\vphantom{#1\rule{0em}{.1em}}}*\cir{l^r};p\save*!L{#1} \restore\save+UC;+UC+<.5em,0em>*!L{\hphantom{#1}}+R **\dir{-} \restore\save+DC;+DC+<.5em,0em>*!L{\hphantom{#1}}+R **\dir{-} \restore\POS+UC+<.5em,0em>*!L{\hphantom{#1}}+R;+DC+<.5em,0em>*!L{\hphantom{#1}}+R **\dir{-} \endxy}}

\begin{document}

\title{No-signaling, entanglement-breaking, and localizability in
  bipartite channels}

\author{Giacomo Mauro D'Ariano}
\affiliation{QUIT group,  Dipartimento di Fisica ``A. Volta'', INFN Sezione di Pavia, via Bassi
  6, 27100 Pavia, Italy.} 
\author{Stefano Facchini} 
\affiliation{QUIT group,  Dipartimento di Fisica ``A. Volta'', INFN Sezione di Pavia, via Bassi
  6, 27100 Pavia, Italy.} 
\author{Paolo Perinotti} 
\affiliation{QUIT group,  Dipartimento di Fisica ``A. Volta'', INFN Sezione di Pavia, via Bassi
  6, 27100 Pavia, Italy.} 
\date{\today}

\begin{abstract}
  A bipartite quantum channel represents the interaction between systems, generally allowing for
  exchange of information. A special class of bipartite channels are the no-signaling ones, which do
  not allow for communication. In Ref. \cite{piani} it has been conjectured that all no-signaling
  channels are mixtures of entanglement-breaking and localizable channels, which require only local
  operations and entanglement. Here we provide the general realization scheme, giving a
  counterexample to the conjecture.
\end{abstract}
\maketitle

Causality is the basic assumption of science, the building block of any mechanism and any prediction
scheme \cite{Pearl}. It is the grey eminence of physical theories, taking apparently different
forms, such as retarded potentials in classical physics, Minkowskian causality in Relativity,
(anti)commutation relations in quantum field theory.  The modern paradigm of causality is
communication, where we identify the causal relation with information exchange. Causality should not
be confused with determinism: indeed, any communication scheme from Alice to Bob can be regarded as
a dependence of the outcome probability distributions at Bob's location on Alice's choice.  It is
easy to recognize that such scheme contains all customary definitions of causality, including
determinism as a very special case.  In synthesis, we define causality as the dependence of a 
probability distribution on a choice. 

In the past quantum entanglement has been claimed as a resource for communication \cite{Herbert},
regarding Alice's choice of local measurement as a way of changing Bob's probabilities---the {\em
  spooky action at a distance} of Einstein \cite{Einstein}.  The impossibility of communicating by
local operations---today commonly referred to as {\em no-signaling}---is instead an immediate
consequence of causality of the theory, as proved in Ref.  \cite{PRURRA}.

In order to have a causal relation between two systems one needs an {\em interaction} between the
two systems $\rA$ (Alice) and $\rB$ (Bob). In Quantum Theory such interaction is represented by a
bipartite channel for $\rA$ and $\rB$, with communication from $\rA$ to $\rB$ corresponding to the
dependence of the local output state of system $\rB$ on the choice of the input state of system
$\rA$. Indeed one can generalize the scheme to the case of $\rA$ and $\rB$ at the input being
different from $\rA'$ and $\rB'$ at the output, considering the causal relation e.~g. from $\rA$ to
$\rB'$. More generally we can include the case of one-dimensional systems, thus recovering also the
situation of mono-partite channels (the case of both inputs and/or both outputs one-dimensional is
uninteresting, since there is no input and /or no output). Notice that causality by definition is a
pairwise relation, whence the bipartite channel is the most general interaction scenario.  For
simplicity we will restrict to finite dimensions, and use the same capital Roman letter to denote
the system and the corresponding Hilbert space, writing $\Lin(\rA)$ for the space of operators on
$\rA$.  The graphical representation of the bipartite quantum channel $\mathcal{C}: \Lin(\rA)
\otimes \Lin(\rB) \rightarrow \Lin(\rA')\otimes \Lin(\rB')$ is given by the following circuit:
\begin{equation}\label{biquantum}
\begin{aligned}
\Qcircuit @C=.8em @R=.8em {
  \ustick{\rA} & \multigate{1}{\mathcal{C}} &\ustick{\rA'}\qw\\
  \ustick{\rB} & \ghost{\mathcal{C}} & \ustick{\rB'}\qw}
\end{aligned}\quad.
\end{equation}
The natural question is now which {\em interactions} allow for communication between input and
output. In Ref. \cite{seminal} it has been shown that not every no-signaling channel is {\em
  localizable}, i.~e. it can be implemented with local operations using entangled ancillas (see
Definition \ref{loc}). In the same reference it has been conjectured that all {\em semi-causal}
channels (namely no-signaling from $\rB$ to $\rA'$, but not necessarily from $\rA$ to $\rB'$) are
also {\em semi-localizable}, namely they are of the form
\begin{equation}\label{circ1}
\begin{aligned}
\Qcircuit @C=.8em @R=.8em {
   \ustick{\rA}&\qw & \multigate{1}{\mathcal{V}_1} &\qw &\qw &\ustick{\rA'} \qw \\
  & & \pureghost{\mathcal{V}_1}    & \ustick{\rE'} \qw&\multigate{1}{\mathcal{V}_2} & \\
  \ustick{\rB}& \qw & \qw &\qw &\ghost{\mathcal{V}_1} &\ustick{\rB'} \qw}
\end{aligned}
\end{equation}
for some system $\rE'$ and suitable quantum channels $\mathcal{V}_1$ and $\mathcal{V}_2$.  Such
conjecture has later been proved in Ref. \cite{semi}. An alternative proof was given in Ref.
\cite{piani}, where the authors also proposed the following
\begin{conjecture}\label{falseconj}
All no-signaling
channels are mixtures of entanglement-breaking and localizable channels.
\end{conjecture}
We will show that Conjecture \ref{falseconj} is false. We will also provide the general realization
scheme for the no-signaling bipartite channel, along with a concrete counterexample to Conjecture
\ref{falseconj}. 

\medskip We will stick on the graphical representation of a bipartite quantum channel in Eq.
(\ref{biquantum}). By ``quantum channel'' we mean a completely positive, trace-preserving map
between the density-matrix space of the input systems and that of the output systems.

The preparation of a state $\rho$ and measurement of a POVM $\{ P_x \}$ on some system $\rA$ are
special classes of channels, graphically represented as
\begin{equation}
\begin{array}{ccc}
\Qcircuit @C=1em @R=1em {
  \prepareC{\rho} &\ustick{\rA}\qw
}
&\quad\quad\quad\quad& 
\Qcircuit @C=1em @R=1em {
  \ustick{\rA} & \measureD{P_x}
}
\end{array}.
\end{equation}
We will use the bijection between states and operators
\begin{equation}\label{Schmidt}
 A = \sum_{mn} A_{mn} \ket{m}\bra{n} \leftrightarrow \Ket{A} = \sum_{mn} A_{mn} \ket{m}\ket{n}
\end{equation}
summarized by the identity 
\begin{equation}
\Ket{A} = (A \otimes I) \Ket{I},
\end{equation}
where $\Ket{I}=\sum_n \ket{n}\ket{n}$ is the (unnormalized) maximally entangled state. It will also
be useful to introduce the Choi-Jamio\l kowski isomorphism between channels
$\mathcal{C}:\Lin(\rA)\rightarrow \Lin(\rB)$ and positive operators on $\rB \otimes \rA$
\begin{equation}
\begin{split}
  & R_\mathcal{C} := \mathcal{C} \otimes \mathcal{I}_\rB
  (\Ket{I}\Bra{I})\\
  & \mathcal{C}(\rho) = \Tr[(I\otimes \rho^T)R_\mathcal{C}],
\end{split}
\end{equation}
where $\rho^T$ denotes the transposition of the operator $\rho$ with respect to the orthonormal
basis in Eq. (\ref{Schmidt}).

We are now in position to make the above mentioned concepts more
precise:

\begin{Def}\label{loc} The channel $\mathcal{C}: \Lin(\rA) \otimes \Lin(\rB)
  \rightarrow \Lin(\rA')\otimes \Lin(\rB')$ is ``localizable'' if it can
  be realized by local operations on $\rA$ and $\rB$ with a shared
  entangled ancilla $\frac1{\sqrt d} \Ket{I}$ on a couple of
  $d$-dimensional systems $\rE_\rA, \rE_\rB$ but without
  communication:
\begin{equation}\label{eq_loc}
\begin{aligned}
\Qcircuit @C=.8em @R=.8em {
  \ustick{\rA} & \multigate{1}{\mathcal{C}} &\ustick{\rA'}\qw\\
  \ustick{\rB} & \ghost{\mathcal{C}} & \ustick{\rB'}\qw}
\end{aligned}
\quad=
\begin{aligned}
\Qcircuit @C=.8em @R=.8em {
    \ustick{\rA}& \qw & \qw &  \multigate{1}{\mathcal{G}_\rA} & \ustick{\rA'}\qw \\
    & \multiprepareC{1}{\frac1{\sqrt d}\Ket{I}} & \ustick{\rE_\rA}\qw  & \ghost{\mathcal{G}_\rA} & \\
    & \pureghost{\frac1{\sqrt d}\Ket{I}} & \ustick{\rE_\rB} \qw & \multigate{1}{\mathcal{G}_\rB} & \\
    \ustick{\rB} &\qw & \qw &\ghost{\mathcal{G}_\rB} & \ustick{\rB'}\qw
}
\end{aligned}\quad .
\end{equation}
\end{Def}

\begin{Def} A bipartite quantum channel $\mathcal{C}:\Lin(\rA)\otimes \Lin(\rB)
  \rightarrow \Lin(\rA')\otimes \Lin(\rB')$ is ``$\rA\nrightarrow \rB'$
    no-signaling'' if $\Tr_{\rA'}[R_\mathcal{C}]= I_\rA \otimes S_{\rB\rB'}$
  where $S_{\rB\rB'}$ is the Choi operator of some channel
  $\mathcal{S}:\Lin(\rB)\rightarrow \Lin(\rB')$.  We say that $\mathcal{C}$ is `` no-signaling'' if
  it is both $\rA\nrightarrow \rB'$ no-signaling and $\rB \nrightarrow \rA'$ no-signaling.
\end{Def}

The following theorem holds

\begin{theorem}\label{realtheo}
  The following are equivalent:
  \begin{enumerate}
    \item The channel $\mathcal{C}:\Lin(\rA)\otimes \Lin(\rB) \rightarrow
  \Lin(\rA') \otimes \Lin(\rB')$ is no-signaling
\item There are equivalent $d$-dimensional quantum systems $\rE_\rA, \rE_\rB$, instruments
  $\{\mathcal{C}_\rA^{(x)}\}_{x\in\rX}$ and $\{\mathcal{D}_\rB^{(x)}\}_{x\in\rX}$ with outcome space
  $\rX$, and channels $\mathcal{C}_\rB^{(x)},\mathcal{D}_\rA^{(x)}$ for each $x\in \rX$ with
\begin{align}
\mathcal{C}_\rA^{(x)}: \Lin(\rA) \otimes \Lin(\rE_\rA) \rightarrow \Lin(\rA')\\
\mathcal{C}_\rB^{(x)}: \Lin(\rB) \otimes \Lin(\rE_\rB) \rightarrow \Lin(\rB')\nonumber\\
\mathcal{D}_\rB^{(x)}: \Lin(\rB) \otimes \Lin(\rE_\rB) \rightarrow \Lin(\rB')\nonumber\\
\mathcal{D}_\rA^{(x)}: \Lin(\rA) \otimes \Lin(\rE_\rA) \rightarrow \Lin(\rA')\nonumber
\end{align}
such that 
\begin{equation}
\begin{split}
\mathcal{C} =&\sum_{x\in X} \mathcal{C}_\rB^{(x)} \circ
\mathcal{C}_\rA^{(x)} (d^{-1}\Ket{I}\Bra{I}_{\rE_A\rE_B})\\ =&\sum_{x\in \rX}\mathcal{D}_\rA^{(x)} \circ
\mathcal{D}_\rB^{(x)} (d^{-1}\Ket{I}\Bra{I}_{\rE_A\rE_B}),
\end{split}
\end{equation}
namely, ${\mathcal C}$ has the two equivalent circuit realizations
\begin{align}
\begin{aligned}\label{finalcirc1}
\Qcircuit @C=.8em @R=.8em {
   \ustick{\rA}& \qw & \qw &  \multigate{1}{\mathcal{C}_\rA^{(x)}} &\qw & \qw  &\ustick{\rA'}\qw \\
& \multiprepareC{1}{\frac1{\sqrt{d}}\Ket{I}}&\ustick{\rE_A}\qw  & \ghost{\mathcal{C}_\rA^{(x)}}&\ustick{\rX}\cw & \pureghost{\mathcal{C}_\rB^{(x)}}\cw & \\
& \pureghost{\frac1{\sqrt{d}}\Ket{I}} & \ustick{\rE_B} \qw&\qw &\qw &\ghost{\mathcal{C}_\rB^{(x)}} & \\
\ustick{\rB} &\qw & \qw&\qw&\qw & \multigate{-2}{\mathcal{C}_\rB^{(x)}} &\ustick{\rB'}\qw
}
\end{aligned}\quad,\\
\nonumber\\
\begin{aligned}\label{finalcirc2}
\Qcircuit @C=.7em @R=.8em {
   \ustick{\rA}& \qw & \qw &  \qw &\qw & \multigate{2}{\mathcal{D}_\rA^{(x)}} &\ustick{\rA'}\qw \\
& \multiprepareC{1}{\frac1{\sqrt{d}}\Ket{I}}&\ustick{\rE_A}\qw  & \qw &\qw & \ghost{\mathcal{D}_\rA^{(x)}} & \\
& \pureghost{\frac1{\sqrt{d}}\Ket{I}} & \ustick{\rE_B} \qw&\ghost{\mathcal{D}_\rB^{(x)}} &\ustick{\rX}\cw &\pureghost{\mathcal{D}_\rA^{(x)}}\cw & \\
\ustick{\rB} &\qw & \qw&\multigate{-1}{\mathcal{D}_\rB^{(x)}}&\qw & \qw &\ustick{\rB'}\qw
}
\end{aligned}\quad.
\end{align}
\end{enumerate}
\end{theorem}

\Proof

Proof of $(1) \Rightarrow (2)$.

$\mathcal{C}$ is $\rB \nrightarrow \rA'$ no-signaling, therefore it can be realized as in Eq.
(\ref{circ1}), where $\rE'$ is a $d'$-dimensional system.  This system can be teleported using the
entangled state $\frac1{\sqrt{d'}}\Ket{I}$ of systems $\rE'_\rA\rE'_\rB$, the Bell measurement
$\Ket{B_x}$ on systems $\rE'$ and $\rE'_\rA$, and classical communication of the outcome $x$ followed
by a controlled unitary $U_x$ on system $\rE'_\rB$, corresponding to the circuit
\begin{equation}
\begin{aligned}\label{telep}
\Qcircuit @C=.8em @R=.8em {
  \ustick{\rA} &\qw &\qw &\qw& \multigate{1}{\mathcal{V}_1}     & \qw                  &\qw &\qw&\ustick{\rA'}\qw \\
  & & & & \pureghost{\mathcal{V}_1}        & \multimeasureD{1}{\Bra{B_x}} &&&\\
  & \multiprepareC{1}{\frac1{\sqrt{d'}}\Ket{I}} &\ustick{\rE'_\rA}\qw &\qw&\qw& \ghost{\Bra{B_x}}            &\pureghost{U^{(x)}}\cw &&\\
  & \pureghost{\frac1{\sqrt{d'}}\Ket{I}}        &\ustick{\rE'_\rB}\qw&\qw&
  \qw& \qw                  & \multigate{-1}{U^{(x)}} & \multigate{1}{\mathcal{V}_2} &\\
  \ustick{\rB} &\qw & \qw&\qw   &\qw               &\qw   &\qw  &\ghost{\mathcal{V}_2} &\ustick{\rB'}\qw
  \gategroup{1}{5}{3}{6}{.7em}{--}
  \gategroup{3}{7}{5}{8}{.7em}{--}
}
\end{aligned}
\end{equation}
(the double wire represents the classical communication of the outcome $x$ of the measurement).

The quantum operation $\mathcal{C}_\rA^{(x)}$ and the channel $\mathcal{C}_\rB^{(x)}$ are the
grouped circuital elements in Eq. (\ref{telep}), and are given by
\begin{align}
  & \mathcal{C}_\rA^{(x)}(\rho) := \Bra{B_x}(\mathcal{V}_1 \otimes
  \mathcal{I}_{\rE'_\rA})(\rho)\Ket{B_x} \\
  & \mathcal{C}_\rB^{(x)}(\rho) := \mathcal{V}_2((U^{(x)} \otimes
  I_\rB)\rho (U^{(x)} \otimes I_\rB)^\dag). \nonumber
\end{align}
The final circuit is thus
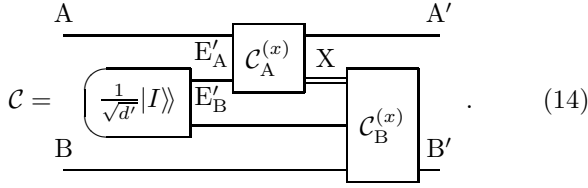
\begin{equation}\mathcal{C}=
\begin{aligned}
\Qcircuit @C=.8em @R=.8em {
   \ustick{\rA}& \qw & \qw &  \multigate{1}{\mathcal{C}_\rA^{(x)}} &\qw & \qw  &\ustick{\rA'}\qw \\
& \multiprepareC{1}{\frac1{\sqrt{d'}}\Ket{I}}&\ustick{\rE'_\rA}\qw  &
\ghost{\mathcal{C}_\rA^{(x)}}&\ustick{\rX}\cw &
\pureghost{\mathcal{C}_\rB^{(x)}}\cw & \\
& \pureghost{\frac1{\sqrt{d'}}\Ket{I}} & \ustick{\rE'_\rB} \qw&\qw &\qw &\ghost{\mathcal{C}_\rB^{(x)}} & \\
\ustick{\rB} &\qw & \qw&\qw&\qw & \multigate{-2}{\mathcal{C}_\rB^{(x)}} &\ustick{\rB'}\qw
}
\end{aligned}\quad.
\end{equation}

Since the channel $\mathcal{C}$ is also $\rA \nrightarrow \rB'$
no-signaling, the same argument gives:
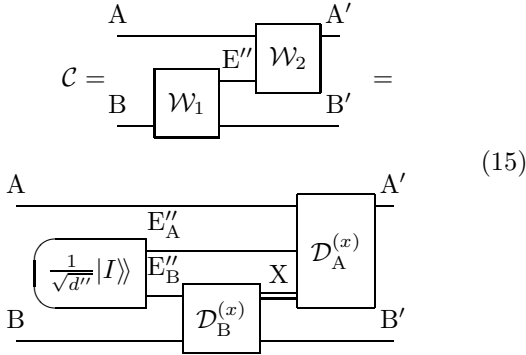
\begin{align}\mathcal{C}=
\begin{aligned}
\Qcircuit @C=.7em @R=.8em {
   \ustick{\rA}&\qw & \qw &\qw &\ghost{\mathcal{W}_2} &\ustick{\rA'} \qw \\
  & & \pureghost{\mathcal{W}_1}    & \ustick{\rE''} \qw&\multigate{-1}{\mathcal{W}_2} & \\
  \ustick{\rB}& \qw & \multigate{-1}{\mathcal{W}_1} &\qw &\qw &\ustick{\rB'} \qw
}\end{aligned}\quad=\nonumber\\ \\
\begin{aligned}\nonumber
\Qcircuit @C=.7em @R=.8em {
   \ustick{\rA}& \qw & \qw &  \qw &\qw & \multigate{2}{\mathcal{D}_\rA^{(x)}} &\ustick{\rA'}\qw \\
& \multiprepareC{1}{\frac1{\sqrt{d''}}\Ket{I}}&\ustick{\rE''_\rA}\qw  & \qw &\qw & \ghost{\mathcal{D}_\rA^{(x)}} & \\
& \pureghost{\frac1{\sqrt{d''}}\Ket{I}} & \ustick{\rE''_\rB} \qw&\ghost{\mathcal{D}_\rB^{(x)}} &\ustick{\rX}\cw &\pureghost{\mathcal{D}_\rB^{(x)}}\cw & \\
\ustick{\rB} &\qw & \qw&\multigate{-1}{\mathcal{D}_\rB^{(x)}}&\qw & \qw &\ustick{\rB'}\qw
}
\end{aligned}
\end{align}
with $\mathcal{D}_\rA^{(x)}$ and $\mathcal{D}_\rB^{(x)}$ given by
\begin{align}
  & \mathcal{D}_\rB^{(x)}(\rho) := \Bra{B_x}(\mathcal{W}_1 \otimes
  \mathcal{I}_{\rE''_\rB})(\rho)\Ket{B_x} \\
  & \mathcal{D}_\rA^{(x)}(\rho) := \mathcal{W}_2((U^{(x)} \otimes
  I_\rA)\rho (U^{(x)} \otimes I_\rA)^\dag). \nonumber
\end{align}

We obtain the statement by defining $\rE_\rA$ and $\rE_\rB$ as
$d$-dimensional systems, where $d:= \max\{d',d''\}$, and embedding
$\rE'_\rJ$ and $\rE''_\rJ$ in $\rE_\rJ$, for $\rJ=\rA,\rB$.

Proof of $(2)\Rightarrow(1)$.

Suppose that $\mathcal{C}$ admits the realization circuit given in Eq.  (\ref{finalcirc1}). We can
group $\rE_\rB$ and $\rX$ in the composite system $\rE'$.  Then $\mathcal{C}$ is also of the form of
Eq. (\ref{circ1}), thus being $\rB \nrightarrow \rA'$ no-signaling, as proved in Ref.
\cite{comblong, semi}.  In the same way, exploiting the second realization circuit in Eq.
(\ref{finalcirc2}), one can prove that $\mathcal{C}$ is also $\rA\nrightarrow \rB'$ no-signaling.
\qed

Theorem \ref{realtheo} shows that the most general no-signaling channel differs from a localizable
channel because it also admits a single round of classical communication, with the constraint that
it must be possible to implement the channel exploiting communication in either directions.

We now provide a counterexample to Conjecture \ref{falseconj}, in terms of a no-signaling channel
that is atomic, (i.~e. it cannot be written as a convex combination of different channels
whence also of no-signaling channels) and that is neither entanglement-breaking nor localizable.
Let $\rA,\rB,\rX_\rA,\rX_\rB,\rW_\rA,\rW_\rB$ be qubits. We define the channel $\mathcal{R}_\alpha$
depending on $\alpha, 0 \leq \alpha \leq 1$:

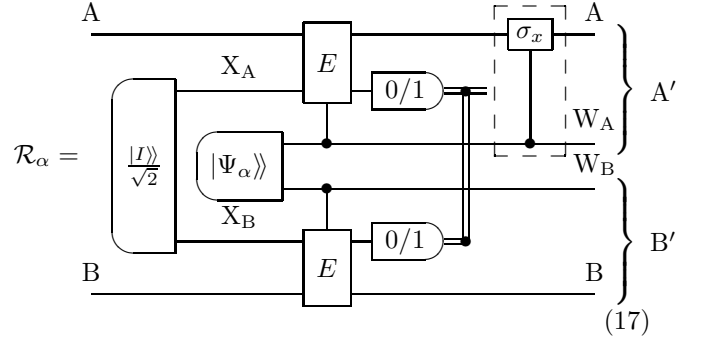
\begin{equation}\label{counter}
\mathcal{R}_\alpha=
\begin{aligned}
\Qcircuit @C=.8em @R=.8em {
  \ustick{\rA} & \qw & \qw & \multigate{1}{E} & \qw&\qw &\qw &\gate{\sigma_x}&\qw & \ustick{\rA}\qw & \\
  & \multiprepareC{3}{\frac{\Ket{I}}{\sqrt2}} & \ustick{\rX_\rA} \qw & \ghost{E} &
  \measureD{0/1}&\control\cw & \cw&  & & & & \rstick{\rA'} \\
  & \pureghost{\frac{\Ket{I}}{\sqrt2}} & \multiprepareC{1}{\Ket{\Psi_\alpha}} & \ctrl{-1} &\qw&\qw\cwx &\qw& \ctrl{-2} &\qw& \ustick{\rW_\rA}\qw&\\
  & \pureghost{\frac{\Ket{I}}{\sqrt2}} & \pureghost{\Ket{\Psi_\alpha}}  & \ctrl{1} &\qw
  &\qw\cwx&\qw&\qw &\qw&\ustick{\rW_\rB}\qw &\\
  & \pureghost{\frac{\Ket{I}}{\sqrt2}} & \ustick{\rX_\rB} \qw & \ghost{E} &\measureD{0/1} &\control\cw\cwx &  & && && \rstick{\rB'}\\
  \ustick{\rB} & \qw & \qw & \multigate{-1}{E} &\qw&\qw&\qw &\qw &\qw&\ustick{\rB}\qw
  \gategroup{1}{11}{3}{11}{.7em}{\}}
  \gategroup{4}{11}{6}{11}{.7em}{\}}
  \gategroup{1}{8}{3}{8}{1em}{--}&
}
\end{aligned}
\end{equation}
where $E$ is the swap operator, $\Ket{\Psi_\alpha} := \sqrt{\alpha} \ket{0}\ket{0} + \sqrt{1-\alpha}
\ket{1}\ket{1}$, the two-qubit gate in the dashed box is a controlled-$\sigma_x$ given by
$\Sigma_{\rA\rW_\rA} := \ket{1}\bra{1}_{\rW_\rA} \otimes (\sigma_x)_\rA + \ket{0}\bra{0}_{\rW_\rA}
\otimes I_\rA$ classically controlled by the outcomes of the measurements on the computational basis
(represented by the circuital element $\Qcircuit @C=.6em @R=.4em { & \measureD{0/1} }$). Notice that
the classical control works as a logical AND, implying that the box $\Sigma_{\rA\rW_\rA}$ is
  performed if and only if both outcomes of the measurements 
$\Qcircuit @C=.6em @R=.4em { & \measureD{0/1} }$ are equal to 1.

\medskip We notice that circuit $\mathcal{R}_\alpha$ in Eq. (\ref{counter}) is implemented using
local operations, entanglement, and one round of classical communication from Bob to Alice, thus
being of the form of Eq.  (\ref{finalcirc2}). One can verify that $\mathcal{R}_\alpha$ can be equivalently
realized applying the controlled-$\sigma_x$ on systems $\rB$ and $\rW_\rB$ as follows
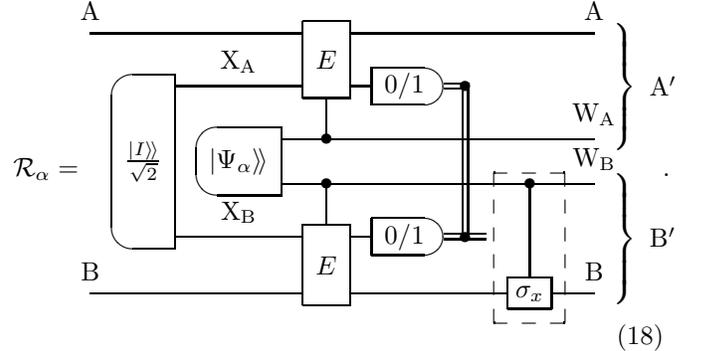
\begin{equation}\label{counter2}
\mathcal{R}_\alpha=
\begin{aligned}
\Qcircuit @C=.8em @R=.8em {
  \ustick{\rA} & \qw & \qw & \multigate{1}{E} & \qw&\qw &\qw &\qw &\qw & \ustick{\rA}\qw & \\
  & \multiprepareC{3}{\frac{\Ket{I}}{\sqrt2}} & \ustick{\rX_\rA} \qw & \ghost{E} &
  \measureD{0/1}&\control\cw & &  & & & & \rstick{\rA'} \\
  & \pureghost{\frac{\Ket{I}}{\sqrt2}} & \multiprepareC{1}{\Ket{\Psi_\alpha}} & \ctrl{-1} &\qw&\qw\cwx &\qw& \qw &\qw& \ustick{\rW_\rA}\qw&\\
  & \pureghost{\frac{\Ket{I}}{\sqrt2}} & \pureghost{\Ket{\Psi_\alpha}}  & \ctrl{1} &\qw
  &\qw\cwx&\qw&\ctrl{2} &\qw&\ustick{\rW_\rB}\qw &\\
  & \pureghost{\frac{\Ket{I}}{\sqrt2}} & \ustick{\rX_\rB} \qw & \ghost{E} &\measureD{0/1} &\control\cw\cwx &\cw  & && && \rstick{\rB'}\\
  \ustick{\rB} & \qw & \qw & \multigate{-1}{E} &\qw&\qw&\qw &\gate{\sigma_x} &\qw&\ustick{\rB}\qw
  \gategroup{1}{11}{3}{11}{.7em}{\}}
  \gategroup{4}{11}{6}{11}{.7em}{\}}
  \gategroup{4}{8}{6}{8}{1em}{--}&
}
\end{aligned}\quad.
\end{equation}
Consequently $\mathcal{R}_\alpha$ also admits a realization of the form given in Eq.
(\ref{finalcirc1}). By Theorem \ref{realtheo}, we can conclude that this is a no-signaling channel.
The Choi-Jamio\l kowski operator of $\mathcal{R}_\alpha$ is:
\begin{equation}\label{eq_choi}
R_\alpha = \sum_{m,n=0}^1 \Ket{K^\alpha_{mn}}\Bra{K^\alpha_{mn}}
\end{equation}
with
\begin{equation}
\Ket{K^\alpha_{mn}}=\left[(\Sigma_{\rA\rW_\rA})^{mn} \otimes \bra{m}_{\rX_\rA}\bra{n}_{\rX_\rB}\right] \Ket{\Phi_\alpha}
\end{equation}
and
\begin{equation}
\Ket{\Phi_\alpha}=(\mathbb{E}\otimes I_{\rA\rB})\left(\Ket{I}_{\rA\rB,\rA\rB}
\otimes \frac1{\sqrt2} \Ket{I}_{\rX_\rA \rX_\rB} \otimes \Ket{\Psi_\alpha}\right),
\end{equation}
where $\mathbb{E}$ denotes the tensor product of the two controlled-swaps.

Using Mathematica, we prove that $\mathcal{R}_{\tilde\alpha}$ with $\tilde \alpha:=1/6$ is a
counterexample by showing that it satisfies the following properties:
(1) It is not entanglement-breaking,
(2) It is not localizable
(3) It is atomic.

Proof of \emph{(1) $\mathcal{R}_{\tilde\alpha}$ is not entanglement breaking.}  A channel is
entanglement breaking if and only if the corresponding Choi-Jamio\l kowski operator is separable.
Thus, we can prove that $\mathcal{R}_{\tilde\alpha}$ is not entanglement breaking by showing that
$R_{\tilde\alpha}$ violates the Peres-Horodecki criterion for separability \cite{ppt1,ppt2}.
According to the criterion, if a state is separable it has a positive definite partial transpose.
Numerically one can check that $R_{\tilde\alpha}$ has a partial transpose with negative eigenvalues,
whence we conclude that it is entangled and $\mathcal{R}_{\tilde\alpha}$ is not
entanglement-breaking.

Proof of \emph{(2) $\mathcal{R}_{\tilde\alpha}$ is not localizable.} If $\mathcal{R}_\alpha$ were
localizable (see Eq. (\ref{eq_loc})), the following observables $A_n, B_m$
\begin{equation}
\begin{aligned}
\Qcircuit @C=.8em @R=.7em {
  &  \prepareC{\ket{n}}  & \multigate{1}{\mathcal{G}_A} & \measureD{\sigma_z}
  && A_n \\
 \multiprepareC{1}{\frac1{\sqrt{d}}\Ket{I}} &\qw& \ghost{\mathcal{G}_A}  &  \\
 \pureghost{\frac1{\sqrt{d}}\Ket{I}} &\qw&\ghost{\mathcal{G}_B} &   \\
  & \prepareC{\ket{m}} & \multigate{-1}{\mathcal{G}_B} & \measureD{\sigma_z} &&
  B_m 
  \gategroup{1}{2}{2}{4}{1em}{--}
  \gategroup{3}{2}{4}{4}{1em}{--}
}
\end{aligned}
\end{equation}
($\Qcircuit @C=.8em @R=.7em { & \measureD{\sigma_z} }$ represents the measurement of $\sigma_z$)
would verify the Cirel'son bound \cite{seminal}:
\begin{equation}
c_\alpha := |\langle A_0 B_0 \rangle + \langle A_0 B_1 \rangle + \langle A_1 B_0
\rangle - \langle A_1 B_1 \rangle| \leq 2\sqrt{2}.
\end{equation}
We have that
\begin{equation}
\langle A_n B_m \rangle = \Tr[(\sigma^z_A \otimes
\ket{n}\bra{n}_A \otimes \sigma^z_B \otimes \ket{m}\bra{m}_B \otimes
I_{W_AW_B}) R_\alpha]
\end{equation}
whence (using expression in Eq. (\ref{eq_choi}) for $R_\alpha$) one finds $c_{\alpha} =
|4-6\alpha|$. Since $c_{\tilde\alpha} = 3 > 2\sqrt{2}$, the Cirel'son bound is violated and
$\mathcal{R}_{\tilde\alpha}$ cannot be localizable.

Proof of \emph{(3) $\mathcal{R}_{\tilde\alpha}$ is extremal.} One can check that the matrices $\{
{K^{\tilde\alpha}}^\dag_{mn}K^{\tilde\alpha}_{m'n'} \} $ are linearly independent.  By Choi's
theorem on extremality \cite{choi} the channel $\mathcal{R}_{\tilde\alpha}$ is extremal.

For a multipartite channel satisfying two different no-signalling
conditions, an analog of Theorem \ref{realtheo} holds. In fact, let us
consider a a channel $\mathcal{C}$ with input systems labelled by a
set of indices $\mathscr{I}$ and output systems labelled by a set
$\mathscr{O}$. Suppose that $\mathcal{C}$ satisfies the following
no-signalling conditions
\begin{equation}\label{multinosign}
\begin{split}
  & \Tr_{\mathscr{O}'}[R_\mathcal{C}] = I_{\mathscr{I}'} \otimes
  S_{\overline{\mathscr{O}'} \cup \overline{\mathscr{I}'}}\\
  & \Tr_{\mathscr{O}''}[R_\mathcal{C}] = I_{\mathscr{I}''} \otimes
  T_{\overline{\mathscr{O}''} \cup \overline{\mathscr{I}''}}\\
\end{split}\quad ,
\end{equation}
for certain subsets $\mathscr{I}', \mathscr{I}'' \subseteq
\mathscr{I}$ and $\mathscr{O}', \mathscr{O}'' \subseteq \mathscr{O}$,
where $\overline{\mathscr S}$ represents the set complement of
$\mathscr S$, and for suitable Choi-Jamio\l kowki operators $S$ and $T$.
Following the proof of Theorem
\ref{realtheo} we can show that two circuits realizing $\mathcal{C}$
are
\begin{equation}
\begin{aligned}
\Qcircuit @C=.6em @R=.6em {
 \ustick{\overline{\mathscr{I}'}} & \qw & \multigate{1}{\mathcal{C}_\rA} &
 \qw&\qw & \ustick{\overline{\mathscr{O}'} } \\
 & \multiprepareC{1}{\Ket{I}} & \ghost{\mathcal{C}_\rA} & \pureghost{\mathcal{C}_\rB}\cw \\
 & \pureghost{\Ket{I}} & \qw &\ghost{\mathcal{C}_\rB} \\
 \ustick{\mathscr{I}' } & \qw &\qw
 &\multigate{-2}{\mathcal{C}_\rB} &\qw
 &\ustick{\mathscr{O}' }
}
\end{aligned}
=
\begin{aligned}
\Qcircuit @C=.6em @R=.6em {
 \ustick{\mathscr{I}'' } & \qw &\qw
 &\multigate{2}{\mathcal{D}_\rA} &\qw
 &\ustick{\mathscr{O}'' }\\
 & \pureghost{\Ket{I}} & \qw &\ghost{\mathcal{D}_\rA} \\
 & \multiprepareC{-1}{\Ket{I}} & \ghost{\mathcal{D}_\rB} & \pureghost{\mathcal{D}_\rA}\cw \\
 \ustick{\overline{\mathscr{I}''}} & \qw & \multigate{-1}{\mathcal{D}_\rB} &
 \qw&\qw & \ustick{\overline{\mathscr{O}''} } \\
}
\end{aligned}\quad .
\end{equation}
In general the subsets $\mathscr{I}', \mathscr{I}''$ are not a
partition of $\mathscr{I}$. In this case we have that the circuits
cannot be realized partitioning the systems between the two local
parties $\rA$ and $\rB$.  In particular the input systems in
$\overline{\mathscr{I}'} \cap \overline{\mathscr{I}''}$ are always
assigned to the party which sends the classical message, and input
systems in $\mathscr{I}' \cap \mathscr{I}''$ are assigned to the party
which receives the classical message (and similarly for output
systems).
One can also consider more complex scenarios,
i.~e. channels with more than two no-signaling conditions of the kind
in Eq. (\ref{multinosign}), or channels with nested conditions, for
example when the Choi-Jamio\l kowski operators $S$ and $T$ in
Eq. (\ref{multinosign}) satisfy no-signaling conditions on their own.
However the analysis of the classical communication required in these
cases is complicated, and is left as on open problem.

In conclusion, we have provided the general realization scheme of
no-signaling channels, giving a counterexample to the conjecture of
Ref. \cite{piani}, stating that such channels are mixtures of
entanglement-breaking and localizable channels. The general
realization scheme looks counter-intuitive, due to the presence of
classical communication. However, the nontrivial constraint is the
fact that an equivalent scheme must exist, with communication in the
reverse direction, and it is remarkable that this constraint is
sufficient to make the channel no-signaling.

The result has an intrinsic foundational relevance, involving the
pivotal role of causality in theoretical physics and computer science.

\par {\em Acknowledgments.---} 
This work is supported by Italian Ministry of Education through grant
PRIN 2008 and the EC through project COQUIT.  We acknowledge
stimulating discussions with Giulio Chiribella, Reinhard Werner,
Vittorio Giovannetti, and Daniel Burgarth.

\end{document}